# Interface induced manipulation of perpendicular exchange bias in Pt/Co/(Pt, Cr)/CoO thin films

N. Akdoğan[1], A. Yağmur[1], M. Öztürk[1], E. Demirci[1], O. Öztürk[1], M. Erkovan*[,2]

[1]Gebze Institute of Technology, Department of Physics, 41400, Kocaeli, Turkey
[2]Sakarya University, Department of Metallurgy and Materials Engineering, 54687, Sakarya, Turkey
*merkovan@sakarya.edu.tr

**Abstract**

Perpendicular exchange bias has been manipulated by changing ferromagnetic film thickness and spacer layer in Pt/Co/(Pt, Cr)/CoO thin films. The exchange bias characteristics, blocking temperature, magnetization of thin films strongly depend on the spacer layer (Pt, Cr) between ferromagnetic and antiferromagnetic layers. While Pt/Co/Pt/CoO thin films show perpendicular exchange bias, Pt/Co/Cr/CoO has easy magnetization axis in the film plane. We have also observed very small hysteretic behavior from the hard axis magnetization curve of Pt/Co/Cr/CoO film. This can be attributed to misalignment of the sample or small perpendicular contribution from Pt/Co bottom interface. We have also investigated the temperature and spacer layer dependent exchange bias properties of the samples by VSM magnetometry. We observed higher $H_{EB}$ and $H_C$ for the thicker Co layer in the Pt/Co/Pt/CoO sample. In addition, onset of exchange bias effect starts at much lower temperatures for Pt/Co/Cr/CoO thin film. This clearly shows that Cr spacer layer not only removes the perpendicular exchange bias, but also reduces the exchange interaction between Co and CoO and thus lowers the $T_B$.

Exchange bias (EB) is shift of the hysteresis loop along the magnetic field axis due to interfacial exchange coupling between ferromagnetic (FM) and antiferromagnetic (AFM) layers after magnetic field cooling the system below the Néel temperature of AFM [1-3]. EB is usually studied in FM/AFM bilayers with the easy magnetization axis in the film plane [4-8]. However, due to demand of ultrahigh capacity in magnetic storage devices, investigation of EB in perpendicularly magnetized systems has recently received much attention [9-15]. In order to observe perpendicular exchange bias (PEB), ultrathin Co layer is one of good candidates since it is known to show perpendicular magnetic anisotropy (PMA) when interact with layers of M=Pt, Pd or Au, due to large interface anisotropy energy of the Co/M interface [16, 17].

In this study, we have investigated the exchange bias properties of Pt(30Å)/**Co(4Å)**/*Pt(5Å)*/**CoO(90Å)**/Pt(30Å) (S1), Pt(30Å)/**Co(5Å)**/*Pt(5Å)*/**CoO(90Å)**/Pt(30Å) (S2) and Pt(30Å)/**Co(4Å)**/*Cr(5Å)*/**CoO(90Å)**/Pt(30Å) (S3) samples grown on Si substrate. We have manipulated PEB in Co/X/CoO thin films by changing spacer layer (X: Pt, Cr) and FM film thickness. The exchange biased Co/X/CoO samples were grown on native oxidized Si wafer under high vacuum conditions at room temperature by magnetron sputtering deposition method. The sample stacks are given in Fig. 1a. The thickness, density and surface/interface roughness of each layer were determined by simulating x-ray reflectivity (XRR) data (Fig. 1b). X-ray photoelectron spectroscopy (XPS) also used to analyze elemental deposition rate and CoO chemical ratio. The details of CoO chemical analysis can be found in Ref. [8].

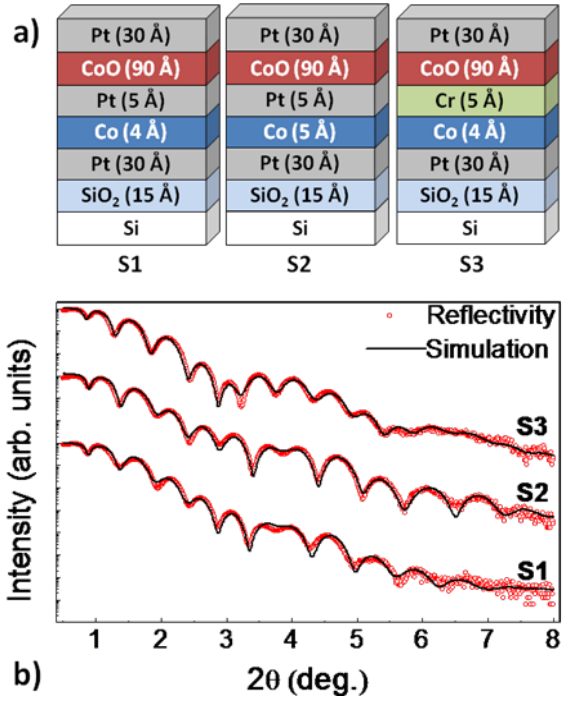

FIG. 1. a) Schematic representation of the samples. b) XRR data and the simulation results. Commercial Rigaku Globalfit program was used for fitting the experimental data.

Magnetic hysteresis curves of the samples were taken by using Vibrating Sample Magnetometry System (Quantum Design PPMS9T). Fig. 2 represents the in-plane and out-of-plane magnetic hysteresis loops measured at room temperature. VSM data reveals that both S1 and S2 have easy magnetization axis perpendicular to the film surface. It is also important to note that S2 has higher remanent magnetization and coercive field compared to S1 in the perpendicular direction. This is explained by increase of Co layer thickness and has a good agreement with previous results [12]. In contrast to the S1 and S2, S3 has easy magnetization axis in the film plane. This occurs just after replacing Pt spacer layer by Cr with the same thickness. We have also observed very small hysteretic behavior from the hard axis magnetization curve of S3. This can be attributed to misalignment of the sample or small perpendicular contribution from Pt/Co bottom interface. The latter needs additional experimental proof by using element specific methods.

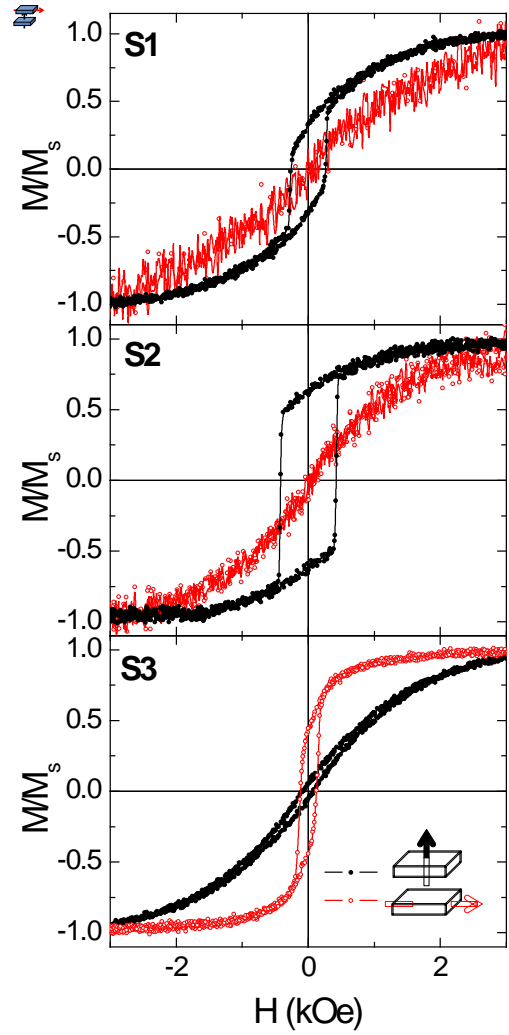

FIG. 2. Hysteresis curves of the samples for both in plane and out of plane geometries taken at room temperature by VSM.

In order to eliminate bulk contributions to the magnetization, we have investigated the samples by using surface sensitive magneto-optical Kerr effect (MOKE) method. Fig. 3 shows hysteresis loops of the samples were taken by SmartMOKE Magnetometry System (Nanosan Instruments) at room temperature for both in-plane (L-MOKE) and out-of–plane (P-MOKE) geometries. In principal MOKE measurements depict similar magnetization mechanism measured by VSM. However, shape of the hysteresis curves drastically changes. Because, VSM collects magnetic information from the whole sample and gives contribution from the bulk. For this reason we observed very rectangular easy hysteresis curve of S1 and S2 by P-MOKE.

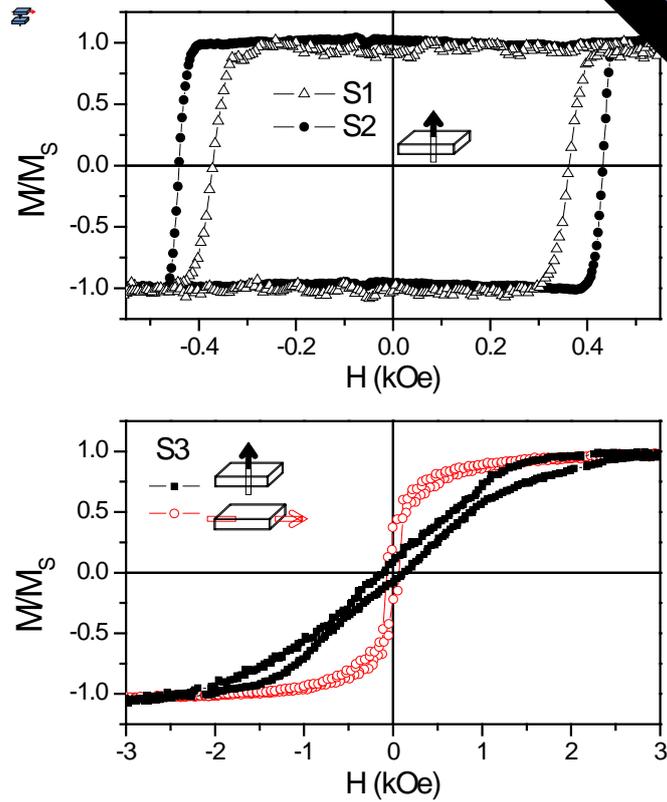

FIG. 3. P-MOKE (out-of-plane) and L-MOKE (in-plane) hysteresis loops of samples measured at room temperature.

Furthermore, we have investigated the temperature and spacer layer dependent exchange bias properties of the samples by VSM magnetometry. Fig. 4 represents coercive ($H_C$) and exchange bias ($H_{EB}$) fields measured at the easy axis of each sample as a function of temperature. In order to eliminate training effects, the samples were heated up to 320 K and then field cooled under 2 kOe down to the target measurement temperature. This heating and recooling procedure was repeated to perform magnetization measurements at each target temperature.

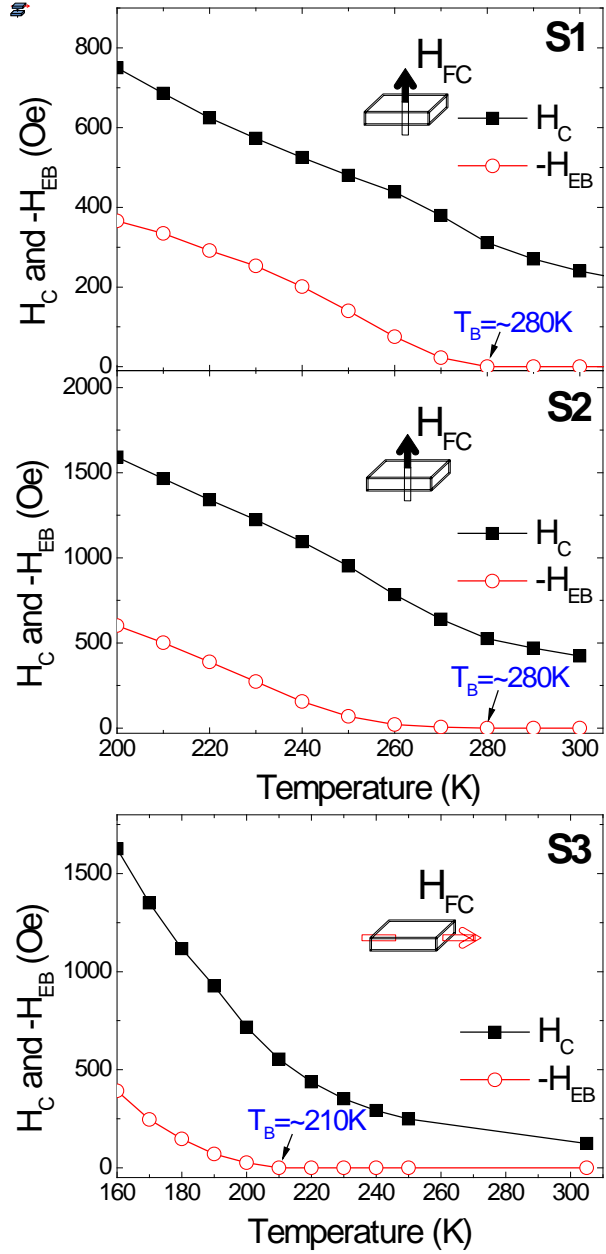

FIG. 4. Temperature dependence of the coercive ($H_C$) and exchange bias ($H_{EB}$) fields of the samples.

It is clear from Fig. 4 that S2 has higher $H_{EB}$ and $H_C$ than S1. This is due to the thicker Co layer in the sample stack. Because, for thicker Co films the system has higher magnetization and $K_{eff}$ value at this thickness limit. This results in higher $H_{EB}$ and $H_C$ as observed in Co/Pt/IrMn system [12]. In addition, S1 and S2 have more or less same blocking temperatures ($T_B$). However, onset of exchange bias effect starts at much lower temperatures for S3. This clearly shows that Cr spacer layer not only removes the perpendicular exchange bias, but also reduces the exchange interaction between Co and CoO and thus lowers the $T_B$.

The significant effect of Cr spacer layer on the exchange bias properties can be explained by the following scenario: The Cr spacer layer reverses the sign of the interface contribution to the total perpendicular magnetic anisotropy of Co. Thus, the negative value of the interfacial anisotropy term implies in-plane behavior and removes perpendicular exchange bias. Also, the impact of interdiffusion at the Co/Cr/CoO interface should not be neglected. The detailed analysis of Co/Cr interface, recently investigated by Reith *et al.*[18], supports our exchange bias results.

In conclusion, we have observed the existence of perpendicular exchange bias for Pt/Co/Pt/CoO thin films. The thicker Co layer enhances the magnitude of exchange bias and coercive field. We have also demonstrated that the type of spacer layer between FM and AFM layers can manipulate the perpendicular exchange bias and blocking temperature of the system. The Cr spacer layer changes the easy magnetization axis from out-of-plane to in-plane, simultaneously reduces exchange bias interaction between Co and CoO, and drastically lowers $T_B$. This is explained by the negative sign of interfacial anisotropy term at the Co/Cr interface of Pt/Co/Cr/CoO thin film.

This work was supported by TÜBİTAK (The Scientific and Technological Research Council of Turkey) through the project numbers 112T857 and 212T217.